\documentclass[aps,prb,onecolumn,superscriptaddress,oneside,floatfix,amsmath,showpacs,amssymb]{revtex4-1}
\usepackage{graphicx} 
\usepackage{dcolumn}
\usepackage{bm}
\usepackage{amssymb,amsmath}
\usepackage{epstopdf}
\usepackage{color}

\begin{document}
\title{Colossal infrared and terahertz magneto-optical activity \\ in a two-dimensional Dirac material}

\author{Ievgeniia O. Nedoliuk}
\affiliation{Department of Quantum Matter Physics, University of Geneva, CH-1211 Geneva 4, Switzerland}

\author{Sheng Hu}
\affiliation{School of Physics \& Astronomy, University of Manchester, Manchester M13 9PL, United Kingdom}
\affiliation{National Graphene Institute, University of Manchester, Manchester M13 9PL, United Kingdom}

\author{Andre K. Geim}
\affiliation{School of Physics \& Astronomy, University of Manchester, Manchester M13 9PL, United Kingdom}
\affiliation{National Graphene Institute, University of Manchester, Manchester M13 9PL, United Kingdom}

\author{Alexey B. Kuzmenko}
\email{Alexey.Kuzmenko@unige.ch}
\affiliation{Department of Quantum Matter Physics, University of Geneva, CH-1211 Geneva 4, Switzerland}


\begin{abstract}
\textbf{When two-dimensional electron gases (2DEGs) are exposed to magnetic field, they resonantly absorb electromagnetic radiation via electronic transitions between Landau levels (LLs) \cite{PoulterPRL91}. In 2DEGs with a Dirac spectrum, such as graphene, theory predicts an exceptionally high infrared magneto-absorption, even at zero doping \cite{ShonJPSJ98,SadowskiPRL06,GusyninJPCM07,AbergelPRB07}. However, the measured LL magneto-optical effects in graphene have been much weaker than expected \cite{SadowskiPRL06,JiangPRL07,OrlitaPRL08,CrasseeNP11,CrasseePRB11,OrlitaPRL11,MaeroPRB14,ChenNC14} because of imperfections in the samples available so far for such experiments. Here we measure magneto-transmission and Faraday rotation in high-mobility encapsulated monolayer graphene using a custom designed setup for magneto-infrared microspectroscopy. Our results show a strongly enhanced magneto-optical activity in the infrared and terahertz ranges characterized by a maximum allowed (50\%) absorption of light, a 100\% magnetic circular dichroism as well as a record high Faraday rotation. Considering that sizeable effects have been already observed at routinely achievable magnetic fields, our findings demonstrate a new potential of magnetic tuning in 2D Dirac materials for long-wavelength optoelectronics and plasmonics.}
\end{abstract}

\maketitle

\begin{figure*}
\includegraphics[width=17cm]{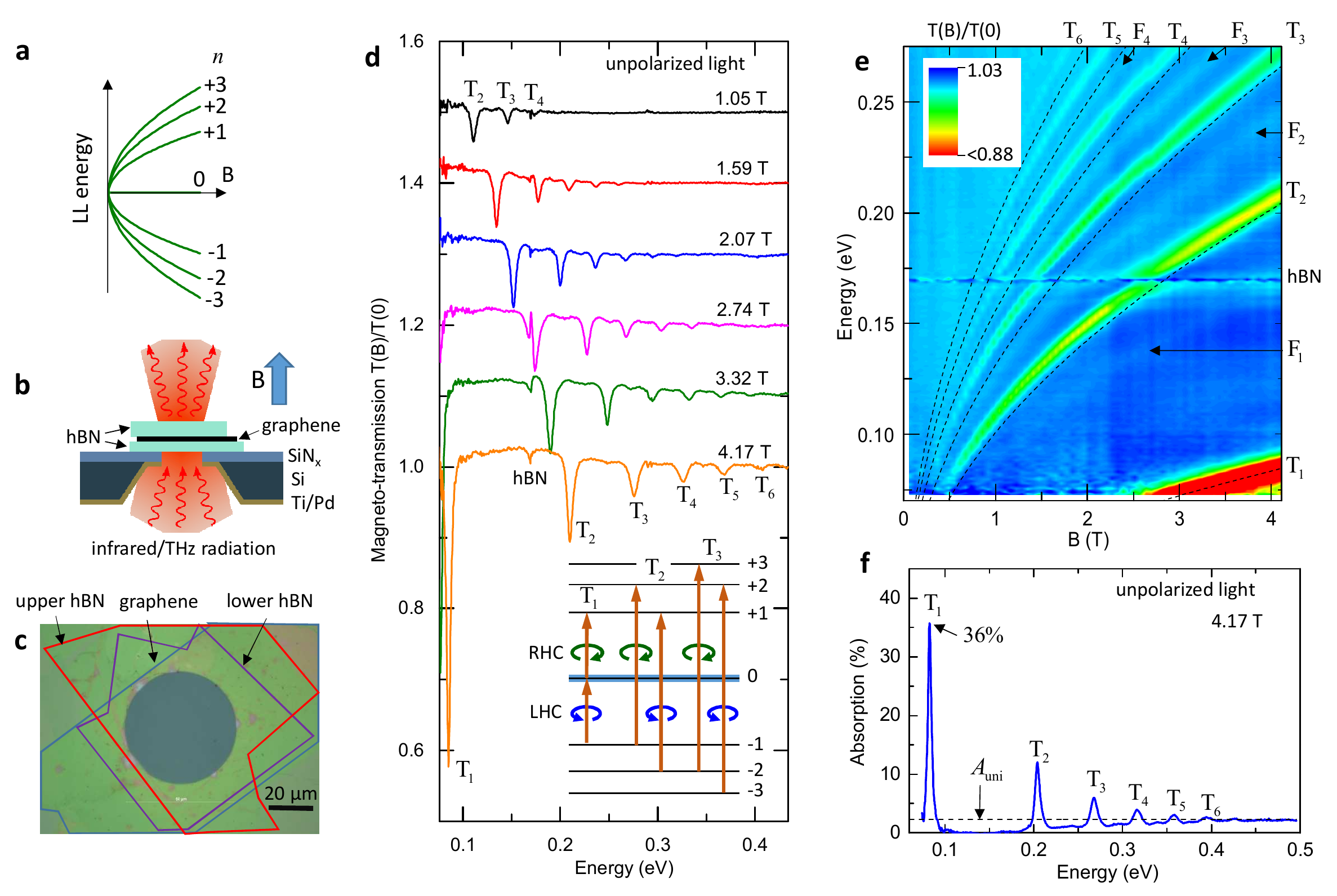}
\caption{\textbf{Interband Landau level transitions in high-mobility encapsulated graphene.} \textbf{a}, Theoretical magnetic field-dependence of the LL energies, $E_{n}(B)=\text{sign}(n)E_{B}\sqrt{|n|}$, where $E_{B} = v_{\text{F}}\sqrt{2e\hbar B}$. \textbf{b}, Schematics of the magneto-infrared transmission experiment. \textbf{c}, Photograph of one of the samples (50 $\mu$m), where the contours of graphene and the hBN layers are indicated by solid lines (other samples are shown in Ref. \onlinecite{Supp}). \textbf{d}, Magneto-transmission spectra $T(B)/T(0)$ at selected magnetic fields in the mid-infrared range in unpolarized light (sample 50 $\mu$m). The inset shows optically allowed transitions $\Delta|n|=\pm 1$ for the RHC and LHC polarizations. The blue line indicates the chemical potential at zero doping. \textbf{e}, Fan diagram of magneto-transmission as a function of the magnetic field and the photon energy (sample 30 $\mu$m). The dashed lines represent theoretical single-electron energies  $T_{n}(B)=E_{B}(\sqrt{n}+\sqrt{n-1})$, where $v_{\text{F}}=1.15\times10^{6}$ m$\cdot$s$^{-1}$. The additional absorption structures $F_{n}$ are due to the forbidden transitions $\Delta|n|=0$ (see inset in Fig. \ref{Figure2}b). \textbf{f}, Absorption spectrum at 4.17 T. The universal absorption $A_{\text{uni}}$ = 2.3\% is indicated by the horizontal line. The feature at 0.17 eV in \textbf{d}, and \textbf{e} is due to a phonon in hBN \cite{Supp}. }
\label{Figure1}
\end{figure*}

In contrast to conventional 2DEGs, the Landau levels in graphene and similar 2D Dirac materials are non-equidistant and non-linear in magnetic field leading to a number of unusual magneto-optical (MO) properties \cite{CastroNetoRMP09}. The LL energies of non-interacting Dirac fermions are given by the formula $E_{n}(B)=\text{sign}(n)E_{B}\sqrt{|n|}$, where $E_{B} = v_{\text{F}}\sqrt{2e\hbar B}$ and $n = 0, \pm 1, \pm 2,...$ is the quantum number, see Fig. \ref{Figure1}a ($B$ is the field intensity, $v_{\text{F}}$ is the Fermi velocity, $e$ is the elementary charge and $\hbar=h/2\pi$ is the reduced Planck constant). The large spacing between the low-order levels is expected to result in unprecedentedly strong optical intensity of the inter-LL transitions \cite{ShonJPSJ98,SadowskiPRL06,GusyninJPCM07,AbergelPRB07}. To the contrary, the so far reported magneto-infrared spectra revealed only a weak, up to a few percent, magnetic increase of the absorption \cite{SadowskiPRL06,JiangPRL07,OrlitaPRL08,CrasseeNP11,CrasseePRB11,OrlitaPRL11,MaeroPRB14,ChenNC14}, possibly due to the electronic and structural imperfections in the large samples used so far. By using a new magneto-infrared microscopy setup we are now able to measure circular polarization-resolved magneto-absorption spectra on small samples of high-mobility graphene encapsulated in hexagonal boron nitride (hBN). We observe a colossal magneto-optical intensity, showing that the reason, or reasons, for the previously seen small magneto-absorption in other types of graphene are not fundamental.

In our experiment, monolayer graphene is encapsulated between two layers of hBN and suspended over a circular aperture in an opaque substrate (Fig. \ref{Figure1}b, \ref{Figure1}c). Due to the small thickness of hBN ($\sim$ 10 nm), it has virtually no influence on the optical properties of the stack, except close to its optical phonon mode at 0.17 eV \cite{Supp}. The magnetic field up to 4.2 T perpendicular to the sample is generated by a compact liquid-He cooled superconducting coil. The base temperature of the sample holder is about 5 K. The suspended part of the membranes may be somewhat warmer due to the external illumination, however we do not observe any dependence of the measured spectra on the radiation intensity, implying the unimportance of the heating and non-linear effects. The intrinsic carrier concentration in as-fabricated devices is below $10^{11}$ cm$^{-2}$ and the mobility is above 10$^{5}$ cm$^{2}$V$^{-1}$s$^{-1}$ \cite{MayorovNL11}. After several thermal cycles the doping grows to $\sim 3\times 10^{11}$ cm$^{-2}$ , likely due to polar adsorbates on hBN (see below). The experiments on three different samples with aperture diameter of 20, 30 and 50 $\mu$m yield consistent results \cite{Supp}.

Fig. \ref{Figure1}d shows mid-infrared magneto-transmission $T(B)/T(0)$ on a fresh undoped sample in unpolarized light as a function of the photon energy $E=\hbar \omega$ at selected field intensities. The interband transitions shown in the inset produce a series of pronounced dips observed up to 0.5 eV at the highest field. Each dip $T_{n}$ comprises two transitions $-n+1\rightarrow n$ and $-n\rightarrow n-1$ active in the right- and left-handed circular (RHC and LHC) polarizations respectively. Our key observation is that the optical transition intensity is extraordinarily high. In particular, at 4.17 T the transmission at $T_{1}$ is reduced by 42\%. For comparison, in unencapsulated graphene epitaxially grown on hBN the transmission reduction at the same field is about 5\% \cite{ChenNC14}, while in multilayer epitaxial graphene on SiC it is below 0.5\% per layer \cite{SadowskiPRL06,CrasseeNP11,CrasseePRB11,OrlitaPRL11,MaeroPRB14}. We conclude that the intensity of the LL absorption depends critically on the graphene morphology and encapsulation.

Fig. \ref{Figure1}e presents the same data as a fan diagram in the coordinates of $B$ and $E$, where the LL transitions are seen as bright traces. For comparison, theoretical single-electron transition energies $T_{n}(B)=E_{B}(\sqrt{n}+\sqrt{n-1})$ are shown (dashed lines), where the Fermi velocity is set (here and later on) to $1.15 \times 10^6$ m$\cdot$s$^{-1}$ in order to match $T_{1}$ at 4 T. With this choice of $v_{\text{F}}$, we see a systematic energy deviation for $n>1$, which means that the Fermi velocity is not the same for different transitions. This is a manifestation of the many-body effects \cite{Supp}, in agreement with previous experiments \cite{JiangPRL07,ChenNC14,FaugerasPRL15,RussellPRL18}.

From this measurement we extract the absorption coefficient $A$, representing the fraction of the incident photons absorbed by graphene \cite{Supp}. Fig. \ref{Figure1}f presents the absorption spectrum at 4.17 T, where one can see a series of sharp peaks culminating well above the universal zero-field value $A_{\text{uni}}\approx2.3\%$ \cite{AbergelPRB07}. The maximum absorption level of 36\% is reached at $T_{1}$, which is comparable to the fundamental limit of 50\% for one-side illumination, achieved when the optical impedance of graphene is twice the impedance of vacuum \cite{Supp}. Interestingly, at the photon energies between the peaks, $A$ is significantly lower than $A_{\text{uni}}$ due to the depletion of the density of states outside the LLs. Specifically, between $T_{1}$ and $T_{2}$ the absorption is below the experimental uncertainty ($\sim$ 1\%). Magnetic tuning therefore offers a huge range of attainable absorption levels in a same sample.

\begin{figure*}
\includegraphics[width=17cm]{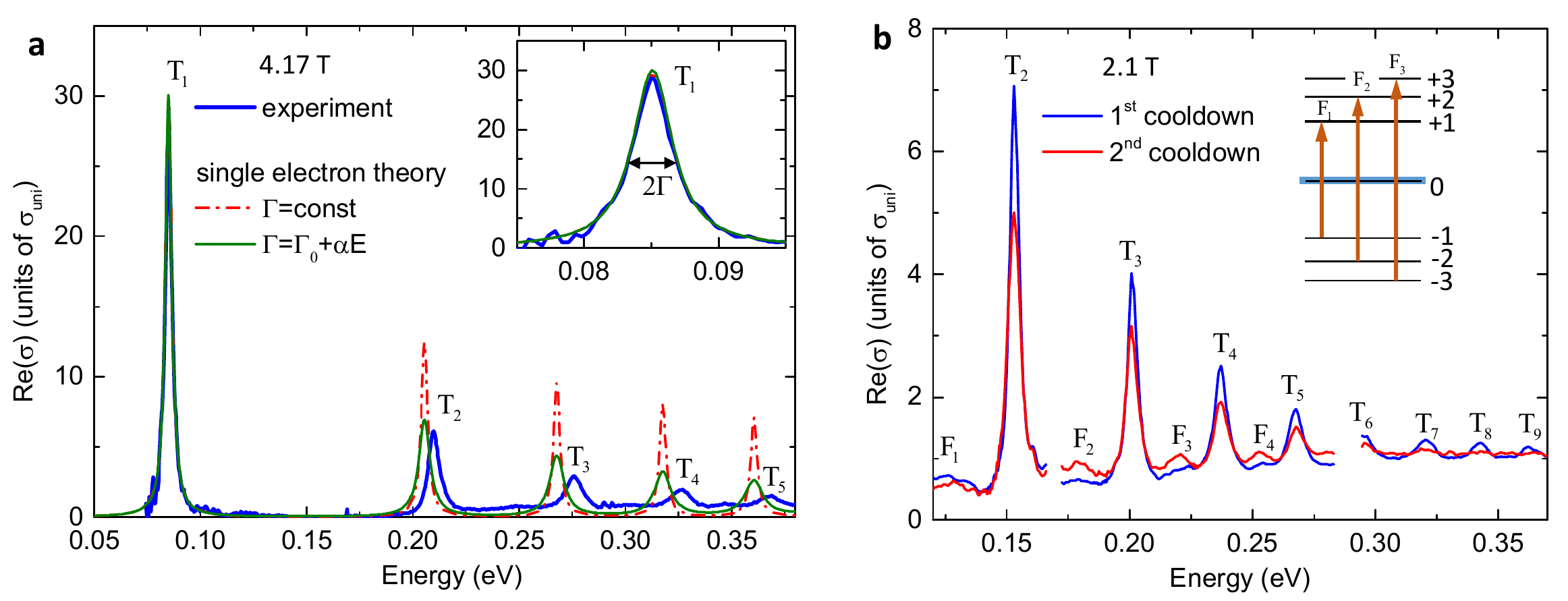}
\caption{\textbf{Analysis of the interband LL transitions.} \textbf{a}, Real part of the optical conductivity $\sigma(\omega)$ at 4.17 T, expressed in the units of $\sigma_{\text{uni}}=e^2/4\hbar$. Blue line: experimental spectrum, red line: single-electron calculation with $v_{\text{F}} = 1.15\times10^{6}$ m$\cdot$s$^{-1}$, $\mu$ = 0, $T$ = 5 K and $\Gamma$ = 1.8 meV, green line: calculation with the same $v_{\text{F}}$, $\mu$ and $T$ but with $\Gamma(E)$ = 0.75 meV + 0.012 $E$. The energy mismatch for the higher-order peaks is due to the many-body effects, as mentioned in the text. \textbf{b}, Optical conductivity at 2.1 T at different cooldowns. The reduction of the allowed transitions $T_{n}$ and the emergence or accentuation of the forbidden transitions $F_{n}$ (inset) in the second run are likely due to imperfections caused by thermal cycling. The breaks at 0.17 eV and 0.29 eV are due to noise introduced by the absorption on the hBN phonon and the atmospheric CO$_{2}$.}
\label{Figure2}
\end{figure*}

To compare these results with theory, we plot in Fig. \ref{Figure2}a the real part of the optical conductivity $\sigma(\omega)$ at 4.17 T (blue curve). On the same graph we present a single-electron calculation (red curve), where each transition is represented by a Lorentzian peak \cite{Supp}. At low temperatures and zero doping the spectral weights, or areas, of the peaks are equal to $W_{n} = \sigma_{\text{uni}}(2E_{B}/\hbar)(\sqrt{n}+\sqrt{n-1})^{-1}$, where $\sigma_{\text{uni}}=e^2/4\hbar$ is the universal optical conductivity \cite{ShonJPSJ98,GusyninJPCM07,AbergelPRB07}. Thus the Fermi velocity fully determines not only the positions but also spectral weights of all peaks. In our first simulation (red curve) we set the half-width at half maximum $\Gamma$ of all Lorentzians to 1.8 meV to meet the observed linewidth of the $T_{1}$ transition. This gives an excellent agreement between the experiment and theory for the first peak (see inset), which means that its spectral weight $W_{1}$ is close to the theoretical value. Interestingly, in this case the maximum conductivity is determined solely by the quality factor $Q=T_{1}/(2\Gamma)$ according to a simple relation \cite{Supp}: $\sigma_{\text{max}} = (4Q/\pi)\sigma_{\text{uni}}$. In the present case $Q\approx 23$ and $\sigma_{\text{max}}\approx 29 \sigma_{\text{uni}}$, exactly as observed. The large quality factor is apparently due to the high electronic mobility in this type of graphene.

The constant-linewidth assumption obviously fails to describe the peak broadening at $n > 1$. Therefore we present another simulation (green curve), where $\Gamma$ grows linearly with the transition energy as suggested earlier \cite{OrlitaPRL11}: $\Gamma(E)= \Gamma_{0}+ \alpha E$. By setting $\Gamma_{0}$ = 0.75 meV and $\alpha = 0.012$ we can match the linewidths of $T_{1}$ and $T_{2}$ simultaneously. While this clearly improves the agreement, the energy-dependent linewidth still does not fully explain our data. In particular, the spectral weight of the high-energy peaks is smaller than the theoretical peak areas. Furthermore, the optical conductivity between the peaks is noticeably higher than the theoretical curve.

To shed more light on this issue, we compare optical conductivity at different cooldowns on the same sample (Fig. \ref{Figure2}b). In the first cooldown (blue curve) the peaks are very pronounced, which allows us to distinguish about ten of them at 2.1 T. In the second run (red curve) they are reduced significantly. Remarkably, this suppression is accompanied by the emergence, or accentuation, of extra peaks $F_{n}$. As seen in Fig. \ref{Figure1}e, the $F$-peaks are interleaved between the $T$-peaks at all magnetic fields, which identifies them as the symmetric interband transitions $-n\rightarrow n$ (inset). The transitions $\Delta |n| = 0$ are optically forbidden (but Raman-active) for an ideal honeycomb lattice. Their presence in our samples may be induced by point defects, strain, random Coulomb potential, electron-electron interactions or a combination of these factors. The variation of the intensity of the forbidden peaks in the same sample allows us to exclude point defects \cite{BriskotPRB13,MaeroPRB14} and interactions as the only reason. On the other hand, thermal cycling may induce strain, while the adsorption of polar molecules on hBN may create a random electrostatic potential. According to the optical sum rule, the infrared activity of the new transitions must incur in any case the intensity reduction of the main peaks, which is indeed observed. We conclude that the LL absorption is highly sensitive to deviations from the ideal graphene structure.

\begin{figure*}
\includegraphics[width=17cm]{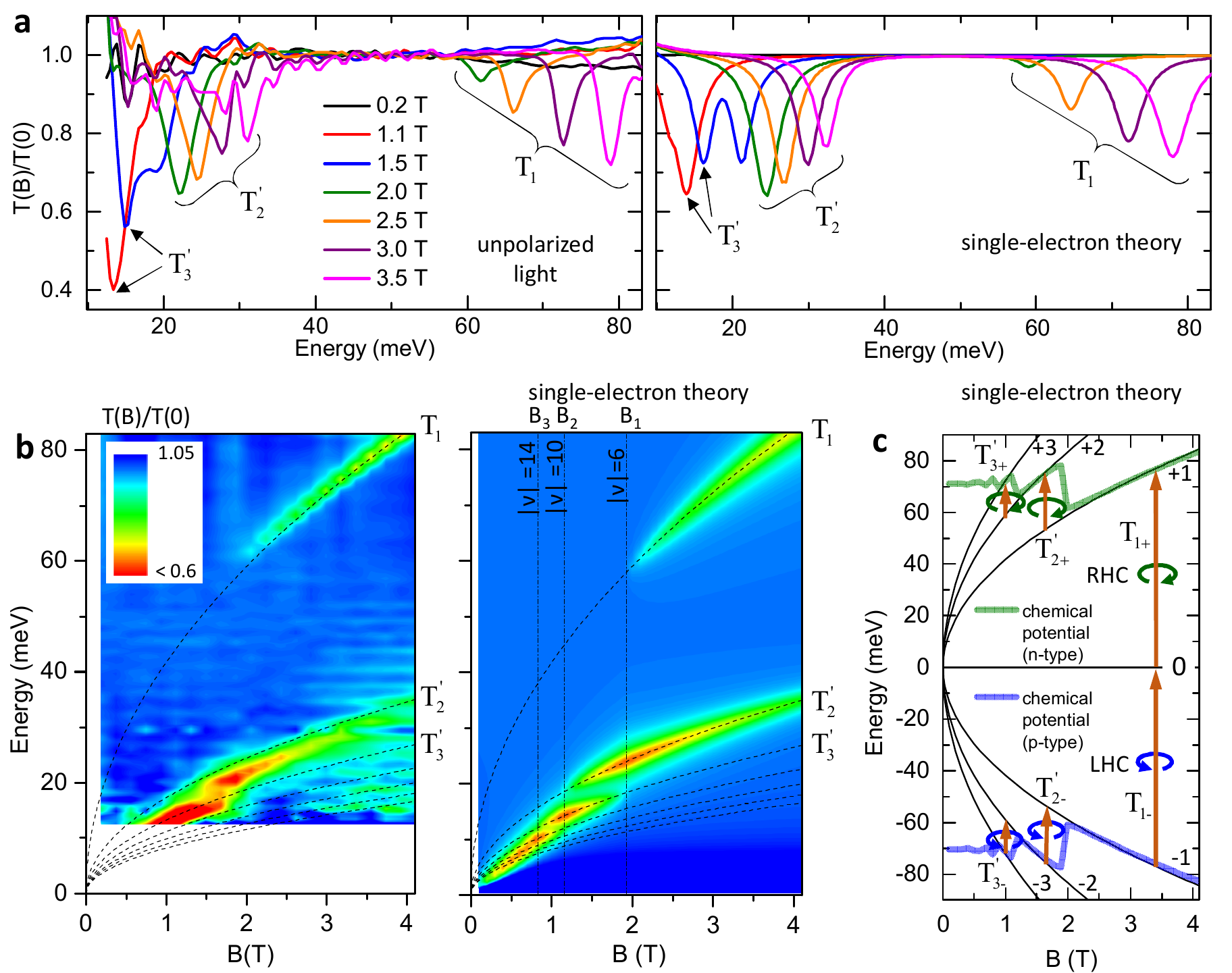}
\caption{\textbf{Far-infrared/terahertz magneto-transmission in weakly doped graphene.} \textbf{a}, Magneto-transmission spectra in unpolarized light at selected values of magnetic field (sample 50 $\mu$m after many thermal cycles). Left panel: experiment.  Right panel: single-electron theory, where the chemical potential is readjusted at every magnetic field in order to conserve the absolute carrier density $|N|=2.8\times 10^{11}$ cm$^{-2}$. \textbf{b}, Magneto-transmission as a function of the photon energy and magnetic field (left: experiment, right: theory). Note that experimental noise grows at the lowest energies as the diffraction limit is approached; the curve for $B$ = 0.2 T (which should be flat in theory) indicates the noise level. \textbf{c}, Calculated magnetic-field dependence of the chemical potential in the cases of p-type and n-type doping (the doping sign can be determined from the Faraday rotation spectra, see Fig.\ref{Figure4}).}
\label{Figure3}
\end{figure*}

Next we focus on the far-infrared/terahertz spectra. The size of our largest sample allows measurement down to 12 meV (3 THz) before reaching the diffraction limit. Fig. \ref{Figure3}a (left) shows unpolarized magneto-transmission at selected fields measured on a sample that underwent several thermal cycles and Fig. \ref{Figure3}b (left) presents the same data as a fan diagram. As the field decreases, the transition $T_{1}$ loses intensity and disappears below $B_{1}\approx 1.9$ T. Additionally, another transition at lower energies ($T'_{2}$) is observed, which gains intensity as the field is lowered and suddenly disappears below $B_{2}\approx1.2$ T. A 1.8 T, a third transition $T'_{3}$ appears and moves beyond the accessible energy range at 1 T. The new structures are due to intraband transitions $-n\rightarrow -n+1$ or $n-1\rightarrow n$, which should have energies $T'_{n}(B)=E_{B}(\sqrt{n}-\sqrt{n-1})$ for non-interacting Dirac fermions. The switching between different branches is expected for doped graphene \cite{OrlitaNJP12}, where the chemical potential $\mu$ jumps between the LLs as the field decreases. Due to the Pauli blocking, a transition $T'_{n}$ disappears at a filling factor $\nu = \pm(4n + 2)$, for the n- and p-type doping respectively. Using the relation $B = hN/(e\nu)$, where $N$ is the carrier concentration, we find that $|N|=6eB_{1}/h \approx 2.8\times 10^{11}$ cm$^{-2}$ (the sign of $N$ will be discussed later). On the right panels of Fig. \ref{Figure3}a and \ref{Figure3}b we present a single-electron calculation, where $\mu$ is adjusted at every magnetic field to match this doping level (Fig. \ref{Figure3}c). This gives an excellent quantitative agreement with the experiment, apart from a small difference between the energy of $T'_{2}$, which might be caused by the many-body effects. As the field intensity is lowered further, the simulation predicts a crossover from the quantum regime to the classical cyclotron resonance with a linear dependence of the cyclotron energy on the magnetic field \cite{OrlitaNJP12}. Remarkably, intraband transitions generate a strong reduction (more than 30\%) of terahertz transmission at rather low carrier concentrations and low magnetic fields ($<$ 1.5 T).

\begin{figure*}
\includegraphics[width=17cm]{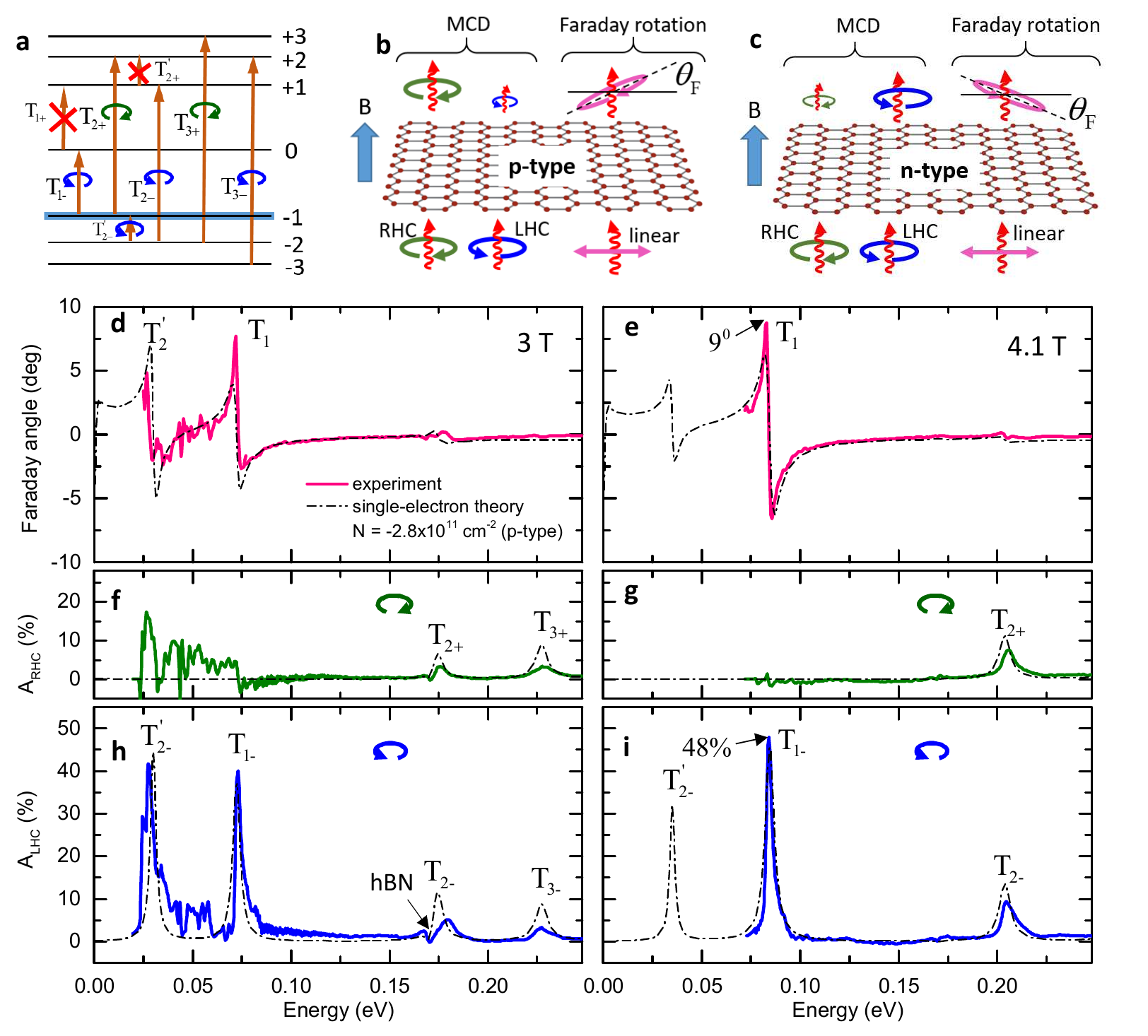}
\caption{\textbf{Faraday rotation and magnetic circular dichroism due to polarization dependent Pauli blocking.} \textbf{a}, Pauli-allowed transitions $\Delta|n|=\pm 1$ at low temperature for the case where the potential is in the LL $n = -1$ ($-6<\nu<-2$). Crosses indicate the Pauli-blocked transitions. \textbf{b} and \textbf{c}, Schematic illustration of FR and MCD for the cases of p- and n-type doping.  \textbf{d}, \textbf{f}, \textbf{h}, Far- and mid-infrared spectra of the Faraday rotation and handedness-resolved absorption  at 3 T (sample 50 $\mu$m). \textbf{e}, \textbf{g} and \textbf{i}, The same data  at 4.1 T (mid-infrared range only). In all panels the single-electron simulation for $N=-2.8\times 10^{11}$ cm$^{-2}$ is shown by dash-dotted lines. Note that the experimental noise grows at the lowest energies as the diffraction limit is approached.}
\label{Figure4}
\end{figure*}

Another important doping-induced phenomenon is a selective Pauli blocking of one of the two oppositely polarized transitions with the same energy (Fig. \ref{Figure4}a). This gives rise to a magnetic circular dichroism (MCD) and a Faraday rotation (FR) with the direction determined by the doping type \cite{CrasseeNP11,CrasseePRB11,Supp,PoumirolNC17} (Figs. \ref{Figure4}b and \ref{Figure4}c). Figs. \ref{Figure4}d and \ref{Figure4}e show the FR spectra at 3 T and 4.1 T respectively. At 3 T, the data are shown in an extended range down to terahertz frequencies. One can clearly see strong spectral structures due to transitions $T_{1}$ and $T'_{2}$. The Faraday angle of 9$^{\circ}$ is observed close to $T_{1}$, which exceeds the maximum values reported so far \cite{CrasseeNP11,PoumirolNC17}.

By combining the Faraday-rotation and magneto-transmission spectra, we can extract the absorption for the RHC and LHC polarizations \cite{Supp,LevalloisRSI15} (Figs. \ref{Figure4}f $-$ \ref{Figure4}i)). We see that transitions $T_{1}$ and $T'_{2}$ are active only in the LHC polarization. This indicates that the doping is p-type, which is typical for environmentally induced charging. The MCD, commonly defined as $(A_{\text{LHC}}-A_{\text{RHC}})/(A_{\text{LHC}}+A_{\text{RHC}})$ is 100\% at these energies. Strikingly, at 4.1 T the LHC absorption is 48\%, nearly matching the ultimate level of 50\% \cite{Supp}. The dash-dotted lines in Figs. \ref{Figure4}d $-$ \ref{Figure4}i present a single-electron calculation for $N=-2.8\times 10^{11}$ cm$^{-2}$ and reproduce all experimental spectra very well for such a simple theory. Interestingly, MCD and FR on higher-order interband transitions are weak, because they are not influenced by the Pauli blocking at this doping level.

The upper theoretical limits for the MO activity in the present case can be easily found within the single-electron approximation. The strongest absorption is produced at $T_{1}$ \cite{Supp}:
\begin{equation}\label{EqAbsMax}
A_{\text{RHC, max}}= \frac{8f_{+}Q\alpha}{(1+4f_{+}Q\alpha)^2},\qquad A_{\text{LHC, max}}= \frac{8f_{-}Q\alpha}{(1+4f_{-}Q\alpha)^2},
\end{equation}
\noindent where $\alpha=e^2/\hbar c$ is the fine-structure constant and $Q$ is the quality factor. The coefficients $f_{+} = f(0)-f(E_{1})$ and $f_{-} = f(-E_{1})-f(0)$ ($f(E)$ is the Fermi-Dirac distribution) describe the effect of the Pauli blocking and are determined by the filling factor $\nu$ \cite{Supp}. The highest possible absorption of $8Q\alpha/(1+4Q\alpha)^2$ is achieved at $\nu=+2$ or -2, where at low temperatures $f_{+} = 1$ and $f_{-} = 0$ or $f_{+} = 0$ and $f_{-} = 1$. This is significantly higher than the peak absorption of $4Q\alpha/(1+2Q\alpha)^2$ at zero doping ($\nu=0$),  where $f_{+}=f_{-}=1/2$  \cite{Supp}. Indeed, in the former case the entire spectral weight of the $T_{1}$ transition is concentrated in one polarization (RHC or LHC), while in the latter case it is shared equally between $T_{1+}$ and $T_{1-}$. This explains why the absorption in Fig. \ref{Figure4}i is much higher than in Fig.\ref{Figure1}f. The maximum FR is achieved at two photon energies slightly detuned from $T_{1}$ \cite{Supp}:
\begin{equation}\label{EqFaraMax}
\theta_{\text{F,max}} = \pm\frac{1}{2}\arctan\frac{2(f_{+}-f_{-})Q\alpha}{\sqrt{(1+4f_{+}Q\alpha)(1+4f_{-}Q\alpha)}}.
\end{equation}
\noindent It vanishes at zero doping and acquires the maximum value of $\pm\arctan(2Q\alpha/\sqrt{1+4Q\alpha})/2$ for $\nu=+2$ or -2. Notably, these relations do not involve the magnetic field explicitly and depend only on the product $Q\alpha$. Therefore, in a non-interacting 2D Dirac system the field intensity affects the MO activity not via the magnetic energy $E_{B}$ but indirectly through the $B$-dependence of the quality factor. It becomes strong if $Q$ is high enough to compensate the smallness of the fine structure constant (1/137). Fortunately, in encapsulated graphene this regime is indeed achieved thanks to a high electronic mobility and small inhomogeneous broadening.

It is worth mentioning that sizeable magneto-absorption and Faraday rotation in the far-infrared range are also observed at low temperatures in heavily doped high-mobility semiconductor quantum wells \cite{PoulterPRL91,IkebePRL10}. As compared to these classical systems, the Dirac 2DEGs offer additional advantages. First, the magneto-absorption can be observed already at zero doping. Second, MCD and FR can be inverted using ambipolar gating \cite{PoumirolNC17}. Third, the magnetic energy grows very fast with the field intensity and the LL peaks fall into the mid-infrared range at reasonably low magnetic fields. Another important consequence of the large LL splitting is the experimentally observed persistence of the $T_{1}$ peak up to room temperature \cite{CrasseePRB11,OrlitaPRL08}.

To summarize, we observe a colossal magneto-optical activity on Landau levels in high-mobility encapsulated graphene in the mid-infrared and terahertz ranges. The key physical ingredients for this phenomenon are the Dirac spectrum and high mobility, both present in graphene/hBN heterostructures. We also observe clear effects of electron-electron interactions, which nevertheless do not prevent us from reaching the maximum electrodynamically allowed magneto-absorption on the transition involving the zeroth LL. A large spread of the attainable absorption levels potentially allows efficient magneto-optical tuning. Importantly, strong MO effects are already observed at low magnetic fields making our findings relevant for applications. The intrinsic MO activity observed here can be enhanced further by modifying optical environment (cavities, Salisbury screen etc.) \cite{FerreiraPRB11,UbrigOE13,JangPRB14} and via plasmonic patterning \cite{JuNN11,PoumirolNC17,TamagnonePRB18}. It would be interesting to do a similar study in graphene/hBN moir\'e superlattices showing the Hofstadter-butterfly physics \cite{HuntScience13,YuNP14} and in twisted bilayer graphene close to the 'magic' rotation angle \cite{CaoNature18}. Our results suggest using graphene and other 2D Dirac materials for magnetically tunable and handedness-sensitive long-wavelength optoelectronics, including magnetic photodetection \cite{KawanoNT13}, modulated quantum cascade lasers \cite{LiangACSP15} and Landau-level based infrared/terahertz emission \cite{MorimotoPRB08,WendlerSR15,WangPRA15}.

\newpage

\end{document}